\documentclass[preprint,5p,times]{elsarticle}

\usepackage{graphicx}               
\usepackage{paralist}        
\usepackage{tabularx}           
\usepackage{comment}
\usepackage{balance}          
\usepackage{multirow}      
\usepackage{fancyvrb}
\usepackage[TABBOTCAP]{subfigure} 
\usepackage{balance}
\usepackage{amsmath}
\usepackage{listings} 
\usepackage{makecell}
\usepackage{booktabs} 
\usepackage[table]{xcolor}
\usepackage{lscape}
\usepackage{xspace}
\usepackage{color, colortbl}
\usepackage[hyphens]{url}
\usepackage{hyperref}
\hypersetup{breaklinks=true}
\usepackage{booktabs}
\usepackage{paralist}
\usepackage{paralist}
\usepackage{color}
\usepackage{url}

\def\BibTeX{{\rm B\kern-.05em{\sc i\kern-.025em b}\kern-.08em
    T\kern-.1667em\lower.7ex\hbox{E}\kern-.125emX}}

\newcommand{\ie}{\emph{i.e.},\xspace}

\newcommand{\eg}{\emph{e.g.},\xspace}

\newcommand{\etal}{\emph{et al.}\xspace}

\journal{Information and Software Technology}

\begin{document}

\newcommand\impr[1]{\textcolor{black}{{#1}}}
\newcommand\rev[1]{\textcolor{black}{{#1}}}
\newcommand\revise[1]{\textcolor{black}{{#1}}}
\newcommand\verify[1]{\textcolor{black}{{#1}}}
\newcommand\major[1]{\textcolor{black}{{#1}}}

\begin{frontmatter}

\title{``\textit{Won't We Fix this Issue?}"  Qualitative Characterization and \\ Automated Identification of Wontfix Issues on GitHub}
\author[ZHAW]{Sebastiano Panichella}
\ead{panc@zhaw.ch}
\author[Sannio]{Gerardo Canfora}
\ead{canfora@unisannio.it}
\author[Sannio]{Andrea Di Sorbo}
\ead{disorbo@unisannio.it}
		\address[ZHAW]{Zurich University of Applied Science, Switzerland}
\address[Sannio]{University of Sannio, Italy}

\begin{abstract}
\textit{Context}: Addressing user requests in the form of bug reports and Github issues represents a crucial task of any successful software project. However, user-submitted issue reports tend to widely differ in their quality, and developers spend a considerable amount of time handling them.\\
\textit{Objective}: By collecting a dataset of around 6,000 issues of \revise{279} GitHub projects, we observe that developers take significant time (i.e., about five months, on average) before labeling an issue as a wontfix.  For this reason, in this paper, we empirically investigate the nature of wontfix issues and methods to facilitate issue management process.\\
\textit{Method}: We first manually analyze a sample of \revise{667} wontfix issues, extracted from heterogeneous projects, investigating the common reasons behind a ``wontfix decision", the main characteristics of wontfix issues and the potential factors that could be connected with the time to close them.  \major{Furthermore,  we experiment with approaches enabling the prediction of wontfix issues by analyzing the titles and descriptions of reported issues when submitted}.  \\
\textit{Results and conclusion}: Our investigation sheds some light on the wontfix issues' characteristics, as well as the potential factors that may affect the time required to make a ``wontfix decision''. \major{Our results also demonstrate that it is possible to perform prediction of wontfix issues} with high average values of \revise{precision, recall, and F-measure (90\%-93\%)}.
\end{abstract}


\end{frontmatter}




\section{Introduction}
\label{sec:intro}

The complexity of modern software systems is growing fast and software developers need to continuously update their source code~\cite{Lehman1980b} to meet users' expectations and market requirements~\cite{SorboPASVCG16}. In this context,   fixing bugs or addressing feature requests and enhancements, reported by users in the form of bug reports~\cite{Anvik:2005:COB:1117696.1117704,AnvikHM06} and Github issues~\cite{BissyandeLJRKT13},  represents a crucial task of any successful software project~\cite{Just:2008:TNG:1549823.1550050,SaloPZ15}. Indeed, during software development and maintenance, issue reports are valuable sources of information for developers interested in improving the quality of the software produced~\cite{AnvikHM06,Azeem:2020}. 

Nevertheless, software changes that are performed to address user-submitted reports often occur under time pressure~\cite{Kim:2006:LDT:1137983.1138027,Usman:2020}, with negative effects on the developers' workloads~\cite{Bertram:2010:CCB:1718918.1718972}. Indeed, user-submitted reports tend to  widely differ in their quality~\cite{Hooimeijer:2007:MBR:1321631.1321639,Bettenburg:2008:MGB:1453101.1453146}, and software developers have to spend a significant amount of time in handling these reports (\eg verifying their content or relevance~\cite{5741332,Tian:2013:DPP:2550526.2550574} and coordinating the teamwork~\cite{Aranda:2009:SLB:1555001.1555045}) for implementing the required changes~\cite{WangZXAS08,BaysalGC09,8961125}.  

In the last decade, research has developed automated solutions to facilitate the issue management and fixing processes, with techniques able to prioritize the requested changes~\cite{Tian:2013:DPP:2550526.2550574,UddinGDNS17}, to detect potential issue misclassifications~\cite{Antoniol:2008:BET:1463788.1463819,HerzigJZ13} and bug duplications~\cite{WangZXAS08}. Hence, most of the proposed tools and prototypes are used to answer critical and relevant questions related to reported issues, \eg ``\textit{Who should fix this bug?}"~\cite{AnvikHM06} or ``\textit{Is It a Bug or an Enhancement?}"~\cite{Antoniol:2008:BET:1463788.1463819}.
However, to the best of our knowledge, only few works investigated the nature of \textit{wontfix} issues, known as ``\textit{bugs that will never be fixed}"~\cite{wontfix}. 
\major{By analyzing more than 6,000 issues from the history of 279 GitHub projects, we observe that developers require time (\ie about five months, on average) before closing an issue with the \textit{wontfix} status.} This means that, in general, developers take about five months to answer the question: ``\textit{Won't We Fix this Issue?}". Starting from this preliminary result, we decided to study the main characteristics of this specific type of issues, thus investigating the main reasons behind a ``\textit{wontfix decision}''.
In addition, we further explored potential factors that could be related to the time to close a general wontfix issue and experimented automated approaches to identify with high accuracy the issues that will be labeled as wontfix, by only analyzing issue titles and descriptions.
To the best of our knowledge, no prior work 
proposed approaches to automatically determine whether an issue will be likely marked (or labeled) as a \textit{wontfix}. 

\impr{Our goal is to support \textit{``community members"}\footnote{\impr{With the term \textit{``community members"} we refer to developers participating in the discussion of the issues, \eg by assigning labels, answering to issues change requests, and identifying also potential issues containing erroneous reports, or requesting features or changes that are out of the project's purposes.}} during the issue management process}. 
\rev{As pointed out by Guo \etal~\cite{Guo:2010:CPB:1806799.1806871}, unfixed bugs receive almost the same amount of attention as fixed bugs: \eg in the Eclipse bug database the average numbers of comments that unfixed bugs and fixed ones receive are  4.5 and 5.8, respectively. Similarly, the average number of comments received by wontfix issues in our dataset is \revise{5.14}.} 
Thus, approaches for timely identifying the issues that will likely be not addressed allow to reduce the unproductive effort (and associated costs) required for triaging and resolving such issues~\cite{wang2020my}. \rev{In particular, early identification of issue success can help (i) project managers allocating resources, (ii) developers focusing their attention on the issues that will be actually addressed, and (iii) customers knowing early if their requirements would be satisfied~\cite{DBLP:journals/jss/Ramirez-MoraOG20}.      
Indeed, the longer the issue that will be likely not fixed remains open, the more it could catch the attention of developers, making them spending efforts in gathering information for attempting to resolve it~\cite{DBLP:conf/csmr/SahaKP14}.} 
Furthermore, being aware of the reasons why developers decide to not fix specific issues could help understanding the software changes that developers consider less relevant. This information could be very useful for improving issue prioritization and triaging mechanisms, in order to better support developers to focus on the issues that will actually get addressed. 

Hence, this paper aims at answering the following research questions: 
\begin{itemize}
\item
\textbf{\textit{RQ$_1$: \impr{What are the main reasons for closing Github issues with the \textit{wontfix} status?}}}
In this research question we qualitatively characterize wontfix issues, by manually analyzing a sample comprising \revise{667} wontfix issues extracted from \revise{97 different projects (developed in C\#) hosted on GitHub}, with the aim of understanding the main reasons behind a ``\textit{wontfix decision}". As initial outcome, we design two different taxonomies. The first taxonomy encompasses the main reasons that pushed users to open issues that later were marked with the \textit{wontfix} label. The second taxonomy models the main motivations given by developers when they decide to close these issues (as wontfix).
\item
\textbf{\textit{RQ$_2$: \impr{What factors relate with the resolution time of \textit{wontfix} issues?}}} 
This research question is a follow-up of the previous one. However, while in {\bf RQ$_1$} we look more at the nature of wontfix issues (\eg investigating the reasons behind a wontfix decision), here we investigate the factors that could be related to the time to close a wontfix issue, observing also whether different wontfix issue types (\ie having different resolution motivations) present different characteristics and, consequently, different resolution times. 
\major{As the time elapsed between the opening and closing of an issue is not necessarily linked to the actual effort spent by developers on the issue itself, RQ$_2$ is also aimed at better characterizing the effort spent on each wontfix issue type.}
\item
\textbf{\textit{RQ$_3$: \impr{Can machine learning be used to automatically identify issues that are likely to be closed as \textit{wontfix}?}}}
In our work we observed that developers take, on average, about five months to figure out that an issue is not worth to be fixed and therefore be labeled as a \textit{wontfix}. In this research question, we want to explore an automated method to anticipate this decision, thus helping developers recognize wontfix issues earlier in the issue management process. During our investigation ({\bf RQ$_1$} and {\bf RQ$_2$}) we observed that most wontfix issues talk about specific aspects (\eg feature enhancements/request, not critical bugs), and that different wontfix issue types tend to experience different resolution times. We conjecture that the topics discussed in the title and the description of an issue report are discriminant and relevant aspects to consider for the fixing of an issue. 
  Based on this consideration we experimented approaches that leverage textual analysis and machine learning techniques to predict  whether an issue  will be marked (or labeled) as wontfix, by analyzing only the titles and  descriptions of reported issues.
\end{itemize}

Results of our study provide insights \revise{into} the nature of wontfix issues, and, in particular, we found that developers mainly tend to close issues (with the \textit{wontfix label}) containing erroneous reports, or requesting features (or changes) that are not relevant or out of the project’s scopes. In addition, the time required to close issues, that developers deliberately decide not to consider, is mainly connected with (i) the issue type, and (ii) the number of participants involved in the related discussions. Finally, our evaluation shows that it is possible to predict whether developers will close an issue as a wontfix by analyzing only the titles and the descriptions of the reported issues and using machine learning and textual analysis techniques. The proposed methodology achieves an average value of \revise{precision, recall, and F-measure ranging between 90\% and 93\%}. \impr{Pragmatically speaking, the high effectiveness  of the experimented ML models in automatically identifying the issues that will likely be closed with the wontfix status could improve the issue management processes allowing developers to focus on more critical  issues.  Indeed, on GitHub, although the use of labels has been envisaged to mark and manage issue reports~\cite{KallisICSME2019}, no automated mechanisms have been provided to support humans during issue labeling.} \major{It is worth pointing out that our work has the only aim of 
providing developers with a tool to more easily identify issues that developers will deliberately avoid to fix by labeling them as \textit{wontfix}}. Other issues may remain open for a long time without any fix~\cite{DBLP:conf/csmr/SahaKP14}, or developers could mark them as ``invalid'', or close them for any other reason~\cite{AnvikHM06}. All these situations are out of the scope of this paper.
  
We believe that our findings not only shed some more light on the nature of wontfix issues, but have the potential to build and/or improve future recommender systems aimed at prioritizing and supporting the issue fixing and the management processes of modern software projects.

In summary, the main contributions of this paper are:
\begin{itemize}
\item Two taxonomies modelling the reasons for opening and closing wontfix issues, along with a manually-labeled dataset (available for replication purposes) of \revise{667} wontfix issues extracted from heterogeneous GitHub projects.
\item Results of our study on the characteristics of the different types of wontfix issues.
\item An automated approach (available for research purposes) able to accurately identify the issues that will likely be not fixed.
\end{itemize}

\begin{figure*}[]
	\centering
	\includegraphics[width=\textwidth]{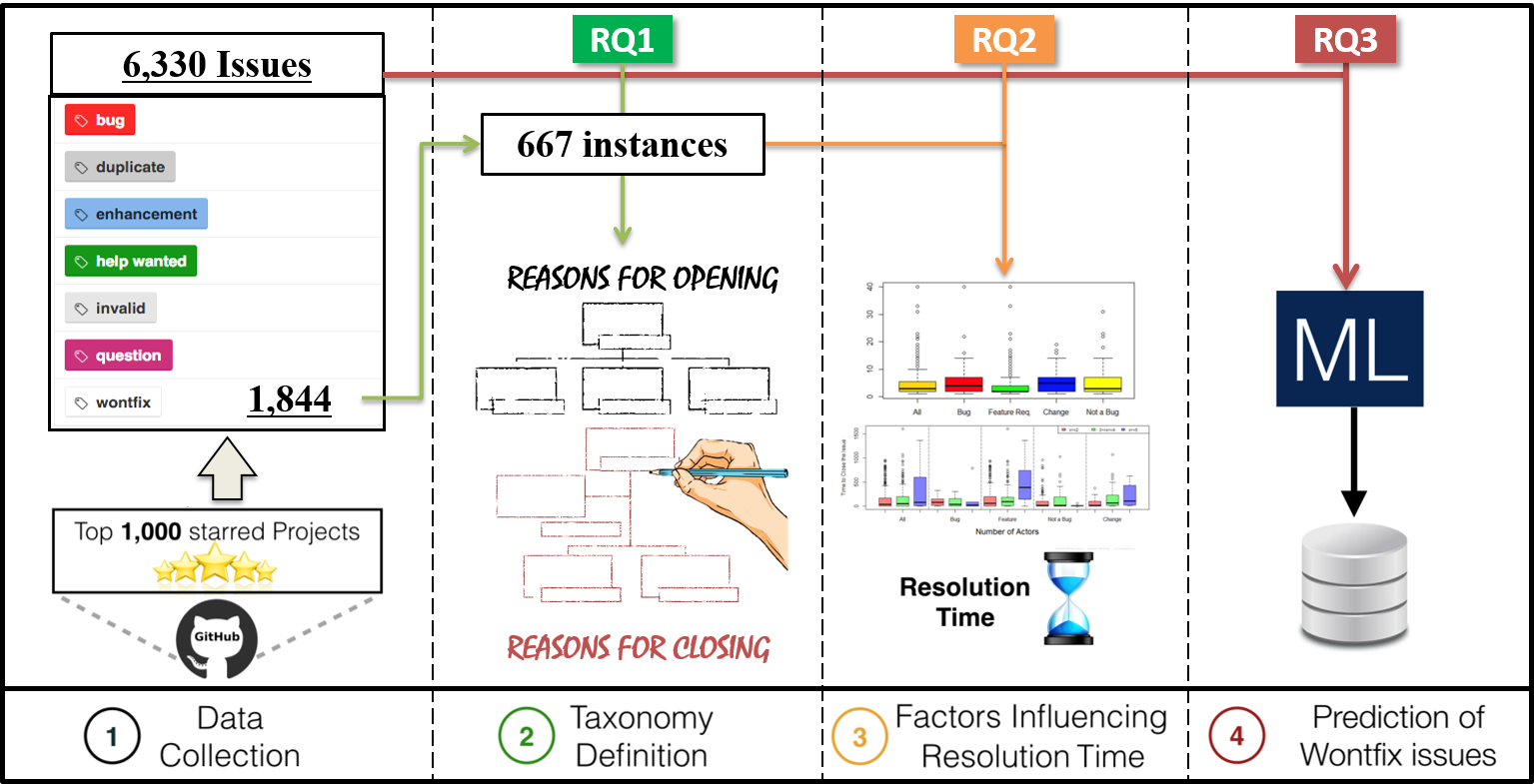}
	\revise{
	\caption{Overview Research Approach}\label{fig:approach}}
\end{figure*}

{\bf Paper structure.} Section \ref{sec:background} discusses the current issue management cycle (with specific reference to GitHub), while Section \ref{sec:related} illustrates the related literature. 
Section \ref{sec:design} details the 
data extraction process and the evaluation methodology adopted to answer our research questions. Section \ref{sec:results} discusses the results achieved in our study, while threats to its validity are reported in Section \ref{sec:threats}. Finally, Section \ref{sec:conclusions} concludes the paper and outlines directions for future work.

\section{Background}
\label{sec:background}
An issue tracking system is a repository where users and team members can submit and discuss issues (\eg bugs and feature requests), ask for advice and share opinions useful for maintenance activities or design decisions~\cite{MurgiaEMSE2018}. GitHub is a social coding platform hosting more than 57 million of repositories\footnote{\url{https://goo.gl/JbS6RE}} which provides advanced version control mechanisms and an integrated issue tracker. 

Any GitHub user can create an issue in a public repository in order to report bugs, require enhancements, or make other kinds of requests. Thus, issues are the primary mean through which GitHub communities collect user feedback. A typical issue on GitHub is described through a \textit{title} and a \textit{description}. 
\major{It is worth noticing that, differently from other bug trackers (\eg JIRA or Bugzilla), the GitHub issue tracker does not provide an explicit description field and the issue description is usually provided in the issue's first comment. Here and in the remainder of the paper, we refer to the first comment of an issue as the \emph{issue description}.}
Moreover, one or more predefined \textit{labels} are used to help in categorizing the issue.
Each issue is assigned to one \textit{assignee} that is responsible for working on it, but \textit{comments} allow anyone  to provide feedback. In order to offer high flexibility, GitHub only provides two issue states (\textit{open} and \textit{closed}), while any other state must be realized via labels. The GitHub issue tracker provides a set of default labels in each repository, including the \textit{wontfix} label which indicates that work will not continue on an issue\footnote{\url{https://help.github.com/articles/about-labels/}}. The \textit{wontfix} label is among the most used labels in GitHub projects~\cite{CabotICR15}.  Although labeling has a positive impact on the effectiveness of issue processing~\cite{LiaoHCFZL18}, the labeling mechanism is scarcely used on GitHub~\cite{CabotICR15}. Thus, automated approaches able to predict the correct labels to assign to issues could stimulate the use of such a mechanism.

In this study we are interested in extracting issues from heterogeneous projects hosted on GitHub, having the \textit{closed} status and the \textit{wontfix} label assigned, in order to investigate common characteristics of this kind of documents. 

\section{Related Work}
\label{sec:related}


\subsection{Issue Management Process and Practices}

Fixing bugs and addressing feature requests or enhancements, reported in the form of bug reports~\cite{Anvik:2005:COB:1117696.1117704,AnvikHM06} and Github issues~\cite{BissyandeLJRKT13}, are crucial tasks for the success of any software project~\cite{Just:2008:TNG:1549823.1550050,SaloPZ15,Panichella15}. For this reason, researchers investigated factors characterizing or affecting the issue management process and practices.

Previous work investigated the aspects that should characterize an informative (or \textit{``good"}) bug report~\cite{Hooimeijer:2007:MBR:1321631.1321639,Bettenburg:2008:MGB:1453101.1453146,Breu:2010:INB:1718918.1718973}. Specifically, Hooimeijer {\em et al.}~\cite{Hooimeijer:2007:MBR:1321631.1321639} presented a first descriptive model of bug report quality, which is based on a statistical analysis of over 27,000 bug reports of  Mozilla Firefox. The evaluation of the model showed its usefulness in reducing the overall cost of software maintenance, suggesting, at the same time, potential features that should be considered when composing bug reports. 
Bettenburg {\em et al.}~\cite{Bettenburg:2008:MGB:1453101.1453146} conducted a survey involving 466 developers and reported that there is a huge information mismatch between what developers need and what users provide in the reported issues. 
Their results suggest that future bug tracking systems should focus on engaging bug reporters, with tools handling bug duplicates. These findings have pushed, in later years, researchers to find solutions to handle the bug duplication problem~\cite{WangZXAS08,Yu:2018:DDP:3196398.3196455}.

Recent research studied socio-technical dynamics~\cite{PanichellaCPO14} concerning the management and fixing of issues~\cite{LiaoHCFZL18} or the handling of pull requests~\cite{Denae2019, Azeem:2020}.  For instance,  
Aranda {\em et al.}~\cite{Aranda:2009:SLB:1555001.1555045} investigated coordination activities around bug fixing tasks by surveying professional developers and found that, even for simple bugs, an inefficient coordination among developers impacts the efficiency of the issue fixing process. These co-ordination problems~\cite{Bertram:2010:CCB:1718918.1718972} are generally  caused by the wrong assignment~\cite{Guo:2011:MBO:1958824.1958887} or re-assignment of bug reports~\cite{BaysalGC09} to developers.  However, in other cases, inefficient bug resolutions are influenced by the length and complexity of issue discussions~\cite{KavalerSHAF17,Destefanis:2018:MAG:3194932.3194936},  the actual knowledge/skills of developers~\cite{RodegheroHDMG16,BissyandeLJRKT13}, and other socio-technical dynamics~\cite{OrtuDKCMT15,OrtuADTMT15,KikasDP15}.  In this context, Breu {\em et al.}~\cite{Breu:2010:INB:1718918.1718973} pointed out that an active  participation of developers represents a crucial aspect for making progress on the reported bugs. For this reason, other work  proposed strategies for determining the appropriate person to assign the reported issues~\cite{AnvikHM06,ZhangYLC16,XuanJRYL17}.

Differently from the aforementioned work, this paper empirically investigates the main characteristics of wontfix bugs and the common reasons behind a wontfix decision.

\subsection{Issue Reports Classification and Prioritization}
\impr{As reported by Cosentino {\em et al.}~\cite{DBLP:journals/access/CosentinoIC17}, even if issues are generally sent to popular projects, the number of pending issues constantly grows~\cite{KikasDP15} and, despite the use of labels has a positive impact on the issue evolution~\cite{CabotICR15}, they are scarcely used by GitHub developers.} Thus, researchers proposed automated solutions to ease the issue management and fixing processes, with techniques that 
leverage well-known methods based on textual analysis~\cite{WangZXAS08}, machine learning~\cite{Antoniol:2008:BET:1463788.1463819,CohenCM04,Bhattacharya:2011:BTP:1985441.1985472}, natural language parsing (NLP)~\cite{BacchelliSDL12,SorboPVPCG15}, and summarization approaches~\cite{MorenoM18,Panichella18,RastkarMM10}
to analyze bug reports information. 
  
Important results in this direction are related to the definition of approaches that automatically classify or analyze the textual content of reported issues~\cite{Zhou:2014:CTM:2705615.2706110}  to derive potential misclassifications~\cite{Antoniol:2008:BET:1463788.1463819,HerzigJZ13}, detecting duplicated bugs~\cite{WangZXAS08} or predicting reopened issues~\cite{Xia:2013:CSS:2495256.2495711}. 
In recent years,  tools have been designed to automatically predict the severity of bug reports~\cite{5741332,ChaturvediS12,LiuWCJ18,Tian:2012:IRB:2420240.2420461}, to support the prioritization  of reported issues~\cite{Tian:2013:DPP:2550526.2550574,UddinGDNS17}, and to estimate the issue life time~\cite{Kim:2006:LDT:1137983.1138027,Kikas:2016:UDC:2901739.2901751,WeissPZZ07,Zhang:2013:PBT:2486788.2486931}.  Finally, to facilitate the process of fixing issues, recent strategies have been proposed to translate bug reports into test cases~\cite{Fazzini:2018:ATB:3213846.3213869}, generating auto-fixes~\cite{6569740}, or recommending relevant classes~\cite{Almhana:2016:RRC:2970276.2970344} for these reports.

In the context of these related studies,  this paper empirically investigates the combination of machine learning and textual analysis techniques to  automatically predict whether issues will be not fixed, by analyzing (only) the titles and  descriptions of reported issues. The closest works to ours are (i) the one by Cabot {\em et al.}~\cite{CabotICR15} who proposed labels to classify issues in open source projects, and (ii) the one by  Guo  {\em et al.}~\cite{Guo:2010:CPB:1806799.1806871} presenting an approach to determine the bugs that will be actually fixed. Finally, recent research~\cite{wang2020my} started investigating the reasons behind wontfix bugs. By manually analyzing a sample of 600 wontfix bugs (extracted from Bugzilla) pertaining to three open-source projects (\ie Eclipse, Mozilla, and OpenOffice), the author identified 12 categories of reasons. Similarly to this research, we perform a manual inspection of wontfix issues. However, we identify, for each inspected issue, both the reason behind the issue opening and the motivation for issue closing, to better understand the co-occurrences between the different sorts of reasons. Besides, we (i) investigate wontfix issues pertaining to 279 heterogeneous projects (hosted on GitHub), identifying further categories of reasons that have not been considered in previous work, (ii) explore the factors that could be related to the time to close a wontfix issue, and (iii) \major{propose an approach able to automatically identify wontfix issues with high effectiveness. To the best of our knowledge, no prior work investigated the nature of wontfix issues on GitHub and proposed approaches to automatically determine whether an issue will be marked as wontfix}.



\section{Study Design}
\label{sec:design}

The \textit{goal} of  our study is to shed some light on the nature of wontfix issues, with the {\em purpose} of building and/or improving recommender systems aimed at prioritizing and supporting the issue fixing and the management processes of modern software projects. Hence, we qualitatively investigate the main reasons behind a ``\textit{wontfix decision"} and explore potential factors that could be correlated with the time to close a general \textit{wontfix} issue. Finally, we experimented with potential strategies to  predict  whether an issue  will be  labeled as a wontfix. Figure \ref{fig:approach} depicts the research approach we followed to answer our research questions.

\subsection{Data Collection}
\label{sec:dataset}

The {\em context} of the study consists of \revise{6,330} issues extracted from the history of \revise{279} open source projects hosted on GitHub, whose characteristics are summarized in Table \ref{tab:systems}.
The selection process we applied for this study is based on a ``criterion sampling"\cite{Patton2002}, according to the following steps:

\begin{enumerate}
 \item \textbf{\revise{Language selection}}: As reported in \cite{DBLP:conf/msr/PadhyeMS14}, projects on GitHub developed in C\# usually have a higher number of external users, and core developers tend to ignore reports from outsiders~\cite{DBLP:conf/oss/HepplerES16, DBLP:conf/oss/DalleBM08}.
 \major{Besides, in a study involving about 100,000 GitHub projects, Bissyand{\'{e}} \etal~\cite{BissyandeLJRKT13} found that GitHub projects developed in C\# usually have higher numbers of issues filed (see Figure 7 in~\cite{BissyandeLJRKT13}). Having a higher number of issues to deal with could increase the likelihood that developers overlook some of these issues due to lack of resources or low priority~\cite{DBLP:journals/tse/FanXLH20}.}
 Thus, in this type of projects \major{(C\# projects)}, we expect to find higher amounts of wontfix issues. For this reason, we selected projects mostly developed through the C\# programming language. 
 \item \textbf{Projects selection:} Recent studies demonstrated that a higher number of issues co-occurs with (i) a higher number of stars received by a GitHub repository\cite{ZeroualiSANER2019}, and (ii) a faster growth of a GitHub project in terms of stars\cite{DBLP:journals/jss/BorgesV18}. 
 Thus, in line with recent empirical studies in software engineering \cite{Nielebock2018, DBLP:journals/infsof/JiangLMFZ17, DBLP:journals/pacmpl/MazinanianKTD17, Hilton:2016:UCB:2970276.2970358}, we selected projects relying on stars information. In particular, in order to consider projects with reliable amounts of issues, the 1000 most popular ones (\ie top starred) have been selected from GitHub. 
 \item \textbf{Metadata extraction}: We collected all closed issues metadata (\eg URL, title, description, resolution date, etc.) from the projects selected according to the aforementioned criteria.   
\end{enumerate}

\impr{The aforementioned steps were performed using the R scripts available in our replication package (under the folder \\ \textit{``1\_Scripts/1\_Data\_Collection}"). In particular, a first script was employed for selecting the projects according to the (1) \revise{language} selection and (2) \revise{projects} selection criteria. This R script addressed the following technical issues: (i) we selected the projects having the higher number of stars; (ii) we selected projects having issues closed, filtering out projects having no issue labels (not all project had issue labels);  (iii) we handled the Github download limits (setting in R a timeout with \\ \textit{``Sys.sleep(40)"}). As result, the script collected the first initial information about the identified projects such as project name, project Github url, project program language, and issue labels. \revise{The other three R scripts we implemented are responsible to collect more detailed information about the issues of the selected projects (e.g., issue url, issue title,  and issue description)}, double-checking that no other issues have been closed as wontfix during the extraction analysis (we added few more wontfix issues with this check). It is worth noticing that during our investigation we observed that specific projects may use custom labels for designating issues that will be not  addressed (\eg \textit{status:wontfix, Resolution-Won't Fix, won't fix, resolved: wontfix, closed:wontfix, wont-fix, Won't Fix, not-fixing, Status-WontFix, WontFix, status: will not fix} and \textit{Cannot fix}). Thus, our scripts consider  issues having the \textit{wontfix} labels described above.}

\revise{Table \ref{tab:systems} reports the number of projects, total number of \textsc{wontfix} and  \textsc{non-wontfix} closed issues mined from these projects. We also report, in the last column of Table \ref{tab:systems}, the median number of issues per project}.

 \textbf{Replication package}. We make available in our replication package\footnote{\url{https://github.com/wontfixRP/wontfixDetection}}  (i) the scripts developed to extract the data used for this research, (ii) all raw data, 
used to generate the data and tables reported in the paper. \rev{In the replication package, we also include the research prototype we used to answer RQ$_3$}.

\begin{table}[t]
\caption{\revise{Characteristics of the analyzed C\# projects.}}
\label{tab:systems}
\scriptsize
\begin{center}
      \resizebox{1\linewidth}{!}{
\revise{
\begin{tabular}{ccccc}
\toprule
 \#Projects    & \#Issues       & \#Wontfix  & \#Non-wontfix  & 	Med.\#Issues.
 \\\toprule
        	           279         &  6,330     	   & 1,844              &  4,486 & 22.69 \\
\bottomrule
\end{tabular}
}
}
\end{center}
\end{table}

\subsection{Analysis Method}
\label{sec:analysis}
\major{In the following, we discuss the research methods used to address each specific research question.}

\subsubsection{RQ$_1$: What are the main reasons for closing Github issues with the \textit{wontfix} status?}
\label{sec:rq1_design}

Answering RQ$_1$ required, as a first step,  to derive a manual labeled \textit{golden set} of wontfix issues, in order to build two taxonomies: (i) a first taxonomy, M$_{opening}$, summarizing the main reasons that pushed users to open issues that later were marked with the \textit{wontfix} label; and (ii) a second taxonomy, M$_{closing}$, modelling the main motivations of developers to close these issues as wontfix.
Therefore, with the aim of manually inspecting a representative sample (99\% confidence level and a \major{margin of error below 4\%}), T$_{sample}$, of the collected wontfix issues (see Section \ref{sec:dataset}), we randomly selected \revise{667} issues from the entire set of wontfix issues. 
These \revise{667} issues belong to 97 projects.  
\impr{Concerning the \revise{667} wontfix issues in our T$_{sample}$, we observed that \revise{286} of them (\revise{42.88\%}) have been opened by end-users (not contributors to the projects), while the remaining \revise{381} issues have been opened by users with different roles in the analyzed projects. In particular, \revise{61} (\revise{16.01\%}) out of these \revise{381} issues  have been opened by repositories’ owners, \revise{225} (\revise{59.06\%}) by organizations’ members, \revise{78} (\revise{20.47\%}) by contributors who had previously committed to the repositories, and \revise{17} (\revise{4.46\%}) by collaborators invited to contribute to the repositories.}


To derive the two taxonomies we used \textit{card sort}, a technique to derive taxonomies from input data~\cite{cardSorting}. We organized card sort in three steps~\cite{DBLP:conf/msr/GuzziBLPD13}: (i) \textit{preparation}, (ii) \textit{execution}, and (iii) \textit{analysis}. \\
\textbf{Preparation}: In this step, we created the cards related to each wontfix issue in T$_{sample}$. Each card represents a wontfix issue and includes: (i) the issue title, (ii) the issue description, (iii) all the messages exchanged in the related discussion, and (iv) all the labels (further to \textit{wontfix}) assigned to the issue by original developers. \\
\textbf{Execution}: Two authors of the paper analyzed the cards applying \textit{open} (\ie without predefined groups) card sort. In particular, the two authors performed an iterative content analysis~\cite{Khalid:2015:IEEE}, starting with two empty lists (one for M$_{opening}$ and the other for M$_{closing}$) of issue categories. Each time they found a new \textit{wontfix issue type} to add to one of two taxonomies, a new category was added to the connected list. The two authors used \textit{pair-sorting}~\cite{DBLP:conf/msr/GuzziBLPD13},  to discuss discrepancies in their thoughts for each card during the card sorting itself and avoid checking the consistency of the sorting and merging the cards in a later phase. \\
\textbf{Analysis}:  To guarantee the integrity of the emerging categories and remove potential redundancies, the two authors performed a second iteration on all the analyzed cards and redefined some of the categories identified in the previous step. Through the card sorting process 22 reasons for M$_{opening}$ and 26 reasons for M$_{closing}$ emerged. During the sorting process we reflected on how they could be further clustered into higher level groups. At the end of this phase, for each taxonomy we identified five high level groups. The resulting taxonomies are described in Section~\ref{sec:rq1_results}, where the set of \revise{667} issues manually validated according to the taxonomies represents our \textit{golden set}.


\subsubsection{RQ$_2$: What factors relate with the resolution time of \textit{wontfix} issues?}
\label{sec:rq2_design}

In order to characterize \textit{wontfix} issues and answer  RQ$_2$, we 
computed the following factors concerning all issues in our \textit{golden set} (\ie All): 

\begin{itemize}
\item \textbf{descriptionLength}: Issue description length (number of characters);
\item \textbf{maxAuthorPercentage}: The proportion of messages posted by the author who posted the majority of messages in the issue discussion;
\item \textbf{majorAuthors}: Number of unique authors who have posted more than one-third  of the overall messages present in the issue discussion;
\item \textbf{meanCommentSize}: Average length of comments (number of characters) in the issue discussion;
\item \textbf{minorAuthors}: Number of unique authors who have posted less than one-third of the overall messages present in the issue discussion;
\item \textbf{nActorsT}: Number of distinct authors participating in the issue discussion;
\item \textbf{nCommentsT}: Number of total comments in the issue discussion;
\item \textbf{timeToCloseIssue}: Time lapse (in days) between issue opening and closing (with the \textit{wontfix} label);
\item \textbf{timeToDiscussIssue}: Time lapse (in days) between issue opening and last comment posted in the issue discussion.
\end{itemize}

These factors allowed us to investigate different issue dimensions, namely (i) the level of participation of the community members to the issue discussion (\textit{nActorsT, maxAuthorPercentage, minorAuthors} and \textit{majorAuthors}), (ii) the discussion's size (\textit{descriptionLength, nCommentsT} and \textit{meanCommentSize}), as well as (iii) timing information about the issue (\textit{timeToCloseIssue} and \textit{timeToDiscussIssue}). 

Moreover, we studied how these factors vary 
when considering the different M$_{closing}$ categories (\ie \textit{Bug, Feature request/enhancement, \major{Not suitable}} and \textit{Change}). 
In particular,  to verify whether statistically significant differences could be observed between the different $M_{closing}$ categories, for each metric $m \in$ \{\textit{descriptionLength,  maxAuthorPercentage,  majorAuthors, meanCommentSize, minorAuthors, nActorsT, nCommentsT, timeToCloseIssue, timeToDiscussIssue}\}, we compared the value distributions obtained for $m$ across the different M$_{closing}$ categories, through the Mann-Whitney U test, a widely used non-parametric test for comparing independent samples \cite{conover1999practical}. \impr{After checking that all the variables of interest (\ie  \textit{descriptionLength, maxAuthorPercentage, majorAuthors, meanCommentSize, minorAuthors, nActorsT, nCommentsT, timeToCloseIssue} and \textit{timeToDiscussIssue}) are not well-modeled by normal distributions (as verified through the Shapiro-Wilk test~\cite{Shapiro1965}), we decided to use a non-parametric test (\ie Mann-Withney), as the assumptions for this test are satisfied by each considered variable.} \major{For coping with multiple comparisons, the Benjamini-Hochberg correction procedure~\cite{benjamini1995controlling} has been adopted to adjust p-values.}

In addition, in order to investigate if some of the considered metrics may influence the time required to close the issue, similarly to the work by Linares-V{\'a}squez \etal~\cite{LinaresFSE2013}, for each metric $m$ we grouped the issues in our \textit{golden set} in different subsets, on the basis of specific values of $m$ (\eg $nActorsT \leq 2$, $3 \leq nActorsT \leq 4$ and $nActorsT \geq 5$) and verified (through the Mann-Whitney U test)  whether statistically significant differences can be observed in the \textit{timeToCloseIssue} distributions obtained for the different subsets. Again, this investigation has been carried out for (i) all the issues in our \textit{golden set} (\ie All), as well as (ii) the various M$_{closing}$ categories.

\subsubsection{RQ$_3$: Can machine learning be used to automatically identify issues that are likely to be closed as \textit{wontfix}?}
\label{sec:rq3_design}

After investigating the nature of wontfix issues, we propose an approach, to automatically predict or classify whether an issue will be labeled as a wontifx. \revise{For achieving this goal, we considered all non-wontfix (4,486) and wontfix (1,844) issues in our dataset (see Table \ref{tab:systems}), collected through the metadata analysis explained in Section~\ref{sec:dataset}}.
Specifically, our approach leverages machine learning (ML) techniques and consists of four main steps:
\begin{enumerate}
\item \textbf{\textit{Preprocessing}}: All terms contained in the titles and descriptions of all 6,330 (4,486 non-wontfix plus 1,844 wontfix) issues in our dataset are used as an information base to build a textual corpus that is preprocessed  applying stop-word removal (using the English Standard Stop-word list) and stemming (\ie English Snowball Stemmer)  to reduce the number of text features for the machine learning techniques~\cite{BaezaYates:1999}.
\revise{In addition, we use the R package \texttt{textclean}, which contains a function called \textit{replace\_html} that allows the automated removal of html tags from the text.}
 The output of this phase corresponds to a Term-by-Document matrix  \textit{M} where each column represents an issue of our dataset and each row represents a term contained in title and or description of the various issues. Thus, each entry \textbf{M$_{[i,j]}$} of the matrix represents the weight (or importance) of the i$-th$ term contained in the j$-th$ issue. 
\item  \textbf{\textit{Textual Feature Weighting}}: Words are weighted using the \textit{tf-idf} score~\cite{BaezaYates:1999}, as opposed to simple frequency counts, because it assigns a higher value to rare words (or group of words) appearing in issues, and a lower value to common ones. This allows identifying the most important words in the issue titles and descriptions. The weighted matrix \textit{M} represents the output of this phase.
  \item  \textbf{\textit{Training and Test sets}}: \revise{For the classification step, we split the matrix related to all issues of our dataset in two parts (50\% each), \ie training and test sets. As training set we considered the sub-matrix, \textbf{M$_{training}$}, obtained by randomly selecting from the original matrix \textit{M} the columns associated to a half of wontfix issues and a half of non-wontfix issues. Vice versa, for the test set we considered the sub-matrix, \textbf{M$_{test}$},  obtained by considering  from the original matrix \textit{M} the columns  associated to the remaining half of wontfix issues and the remaining half of non-wontfix issues}. 
\item \textbf{\textit{Classification}}: We automatically classify wontfix issues in the test set by relying on the output data obtained from the previous step consisting of the matrix \textbf{M$_{training}$} and  \textbf{M$_{test}$}. Specifically, to increase the generalizability of our findings, we experimented (relying on the Weka tool \footnote{\url{https://www.cs.waikato.ac.nz/ml/weka/}}) different machine learning techniques, namely, the standard probabilistic Naive Bayes classifier, the sequential minimal optimization (SMO) algorithm\footnote{\url{https://goo.gl/BwMjzS}}, and the J48 tree model. It is important to note that, the choice of these techniques is not random since they were successfully used for bug reports~\cite{Antoniol:2008:BET:1463788.1463819,ZhouTGG16} or vulnerability~\cite{DBLP:journals/jss/RussoSVC19} classification, recent work on user reviews analysis~\cite{PanichellaSGVCG15,SorboPASVCG16,SorboGVP21}, and several works on bug prediction~\cite{BasiliBM96,ZimmermannN09}. 
\end{enumerate}

\impr{It is worth highlighting that, as stated in Section~\ref{sec:intro}, for predicting whether an issue will be labeled as \textit{wontfix}, the machine learning models have been trained  by exclusively using information that are immediately available at the issue opening (\ie issue title and description, without considering the other features), this to simulate a more realistic scenario in which the automated classification can really help developers identifying the issues that will  likely be not fixed.}

\major{To evaluate the performance of the experimented ML techniques, we adopted well-known information retrieval metrics, namely  precision, recall, and F-measure~\cite{BaezaYates:1999}. 
It is important to specify that, as described in the steps 3 and 4 of our approach, we apply a cross-projects setting to train the ML models on data coming from different projects}. This choice was made to ensure that a more general classification model is trained. To complement the evaluation process and alleviate concerns related to overfitting and selection bias, we also provide 
the classification results of the experimented machine learning models, by computing a 10-fold validation strategy.

\section{Results}
\label{sec:results}
This section discusses the results of our empirical study.

\subsection{\textbf{RQ$_1$: Reasons for Wontfix Issues}}
\label{sec:rq1_results}
\begin{table}[t!]
   \scriptsize
   \begin{center}\impr{
       \caption{\revise{Motivations for issue opening (only wontfix issues)}}
       \label{tab:label2}
       \begin{tabular}{c 
           m{3.5cm} c} \toprule
	     \textbf{Category} 
	     & \textbf{Motivation} & \textbf{Freq.}  \\ \toprule
\multirow{5}{*}{Functional Aspects} 
& Feature enhancement & \revise{42.6\%} \\
 & Reported a bug & \revise{29.1\%} \\
 & Feature Request & \revise{24.3\%} \\
 & Unexpected/unclear functionality behaviour & \revise{1.0\%} \\
 & Unknown crash & 0.1\% \\ \midrule
\multirow{6}{*}{Problem} 
& Performance Issue & \revise{2.1\%} \\
 & Testing related issue & \revise{1.2\%} \\
 & Security issue & \revise{0.6\%} \\
  & Compilation issue  & \revise{0.4\%} \\
 & Browser issue & \revise{0.3\%} \\
 & Persistence issue & \revise{0.3\%} \\ \midrule
\multirow{3}{*}{Configuration}
 & Tool version no longer supported & \revise{0.6\%} \\
 & Configuration Problem & 0.1\% \\
 & Backup problems & 0.1\% \\ \midrule
\multirow{3}{*}{Documentation} 
 & Question & 2.8\% \\
 & Request of adding an example in the documentation & 0.6\% \\
 & Not clear or incorrect code/software examples & 0.1\% \\ \midrule
\multirow{5}{*}{Other} 
 & Development aspects & \revise{1.3\%} \\
  & Comment on the software (Not a real issue post) & \revise{0.9\%} \\
 & Improvement of graphical aspects & \revise{0.4\%} \\
 & Current feature beyond the scope of the library & \revise{0.1\%} \\
 & General comment & 0.1\% \\ \bottomrule
 	\end{tabular}}
  \end{center} 
\end{table}

\begin{table}[h!]
   \scriptsize
   \begin{center}\impr{
       \caption{\revise{Motivations for issue closing (only wontfix issues)}}
       \label{tab:label1}
       \begin{tabular}{m{2cm} m{4cm} c} \toprule
	     \textbf{Category} & \textbf{Motivation} & \textbf{Freq.}  \\ \toprule
   \multirow{6}{*}{\shortstack{Feature request/\\enhancement}} 
   & Feature request/enhancement already implemented or not needed & \revise{47.8\%} \\
   &  Not relevant request & \revise{4.5\%} \\
   & It was fixed in the context of previous bug fixes & \revise{0.7\%} \\
   & Too expensive feature request & 0.3\% \\
   & Feature request that will be implemented in the near future & \revise{0.1\%} \\
   & Already Implemented feature request by an external contributor of the project & 0.1\% \\ \midrule
    \multirow{3}{*}{Change} 
   & Not relevant change & \revise{14.5\%} \\
   & No time to work on this change & \revise{1.0\%} \\
   & Requested change leading to further problems & \revise{0.7\%} \\ \midrule
   \multirow{8}{*}{\major{Not suitable}} 
   & Not a bug & \revise{10.8\%} \\
   & Configuration/backup problem on the user side & \revise{3.9\%} \\
   & Duplicated issue & \revise{2.4\%} \\
   & Unclear wrong/usage of a functionality & \revise{1.6\%} \\
   & Problem already fixed with the new version & \revise{1.3\%} \\
   & Problem fixed updating the new version of a dependent library/tool & \revise{0.7\%} \\
   & Tool version no longer supported & \revise{0.4\%} \\
   & Not replicable bug & 0.1\% \\ \midrule
   \multirow{5}{*}{\shortstack{Bug}} 
   & Impossible to fix the issue or too expensive change & \revise{4.9\%} \\
   & Not a critical bug & \revise{2.5\%} \\
   & It will be fixed in future & \revise{1.8\%} \\
   & Difficult to fix or to replicate  & \revise{1.6\%} \\
   & Unknown crash & 0.1\% \\ \midrule
   \multirow{4}{*}{Other} 
   & General comment from a user & 0.4\% \\ 
   & Closed by the user & 0.1\% \\
   & It was a test failure & 0.1\% \\
   & Updated the documentation on wiki & 0.1\% \\ \bottomrule
	\end{tabular}}
  \end{center}
\end{table}

To explore the common motivations for closing issue reports which developers deliberately  avoid to consider/fix (\ie \textit{wontfix}), and answer RQ$_1$, we performed a manual analysis of a sample of issues extracted from our data collection (as described in Section \ref{sec:rq1_design}). More specifically, such \textit{golden set} encompasses \revise{667} closed issues (with the \textit{wontfix} label) extracted from 97 distinct projects hosted on GitHub.



Each issue in the sample has been marked with two labels: (i) the motivation behind the issue opening, as stated by the issue reporter (\ie the motivation for issue opening, M$_{opening}$), and (ii) the reason for its closing (with the \textit{wontfix} status), as declared by developers within the issue discussion (\ie the motivation for issue closing, M$_{closing}$). 

For M$_{opening}$, we found 22 different motivations (reported in Table \ref{tab:label2} along with their frequencies within the analyzed sample), that have been grouped  in five distinct categories. 
It is worth to highlight that \revise{64} (\revise{9.6\%}) issues have been assigned to more than one M$_{opening}$ category, since they have been opened for multiple purposes (this explains why the sum of percentage values in Table \ref{tab:label2} 
is higher than 100\%). For M$_{closing}$, we found 26 distinct motivations (reported in Table \ref{tab:label1}). Such motivations have been clustered in five categories. 
Only \revise{23} issues (\revise{3.4\%}) have been marked with multiple M$_{closing}$ motivations, this is mainly due to the fact that community members usually tend to provide a precise indication for not fixing an issue.  

In most cases, the reasons for opening an issue are related to bugs reporting, feature requests or enhancements, and only in few cases, by other requests (\eg clarification questions, performance, and testing related aspects). As expected, the majority (\ie \revise{648}, \revise{97.15\%}) of issues in our sample have been opened in order to signal troubles dealing with functional aspects (see Table \ref{tab:label2}). As illustrated in Table \ref{tab:label2}, many of the issues belonging to the \textit{Functional Aspects} category are opened in order to (i) request improvements for specific features (\ie \textit{Feature enhancement}, \revise{42.6\%}), (ii) report a bug (\ie \textit{Reported a bug}, \revise{29.1\%}), or (iii) require a new feature to be implemented (\ie \textit{Feature request}, \revise{24.3\%}). 
As anticipated, requests for fixing defects not strictly related with the software features (\ie \textit{Problem} in Table \ref{tab:label2}), the requests about software documentation  (\ie \textit{Documentation} in Table \ref{tab:label2}), and configuration problems (\ie \textit{Configuration} in Table \ref{tab:label2}) resulted in rare motivations for opening issues.

Community members usually decide to ignore issues (i) containing erroneous reports, or (ii) requesting features or changes that are out of the project's purposes (\eg requests of improving the performance or GUI associated to a feature, or adding a functionality that is already present in the system). \major{As a matter of fact, 319 (47.8\%) issues in our sample, have been closed with the motivation that the requested features/enhancement were not needed or had been already implemented, while 142 (21.3\%) issues erroneously reported problems, which have been proved to be not suitable (see Table \ref{tab:label1})}. 
\impr{Indeed, only \revise{63} (\revise{9.4\%}) issues revealed actual bugs, which developers decided to not fix,
as they have been often evaluated as (i) too expensive to fix (\ie \textit{Impossible to fix the issue or too expensive change}, 52.4\% of issues signaling actual bugs), (ii) not critical (\ie \textit{Not a critical bug}, \revise{27\%} of issues signaling actual bugs), or (iii) that will be fixed in the future (\ie \textit{It will be fixed in future}, \revise{19}\% of issues signaling actual bugs).}
In addition, issues reporting change requests (\ie \textit{Change} in Table \ref{tab:label1}) are mainly closed (\ie \revise{14.5\%} of issues) because the change they propose are judged as not strategically relevant by community members. \major{We argue that results in Table \ref{tab:label1} could be useful to implement more accurate analysis (not necessarily based on binary classification) for future work, such as multi-label issue classification \cite{TanZS20} and issue prioritization \cite{DhasadeVC20}}.

\begin{figure*}[t!]
\centering 
	\includegraphics[width=0.7\textwidth]{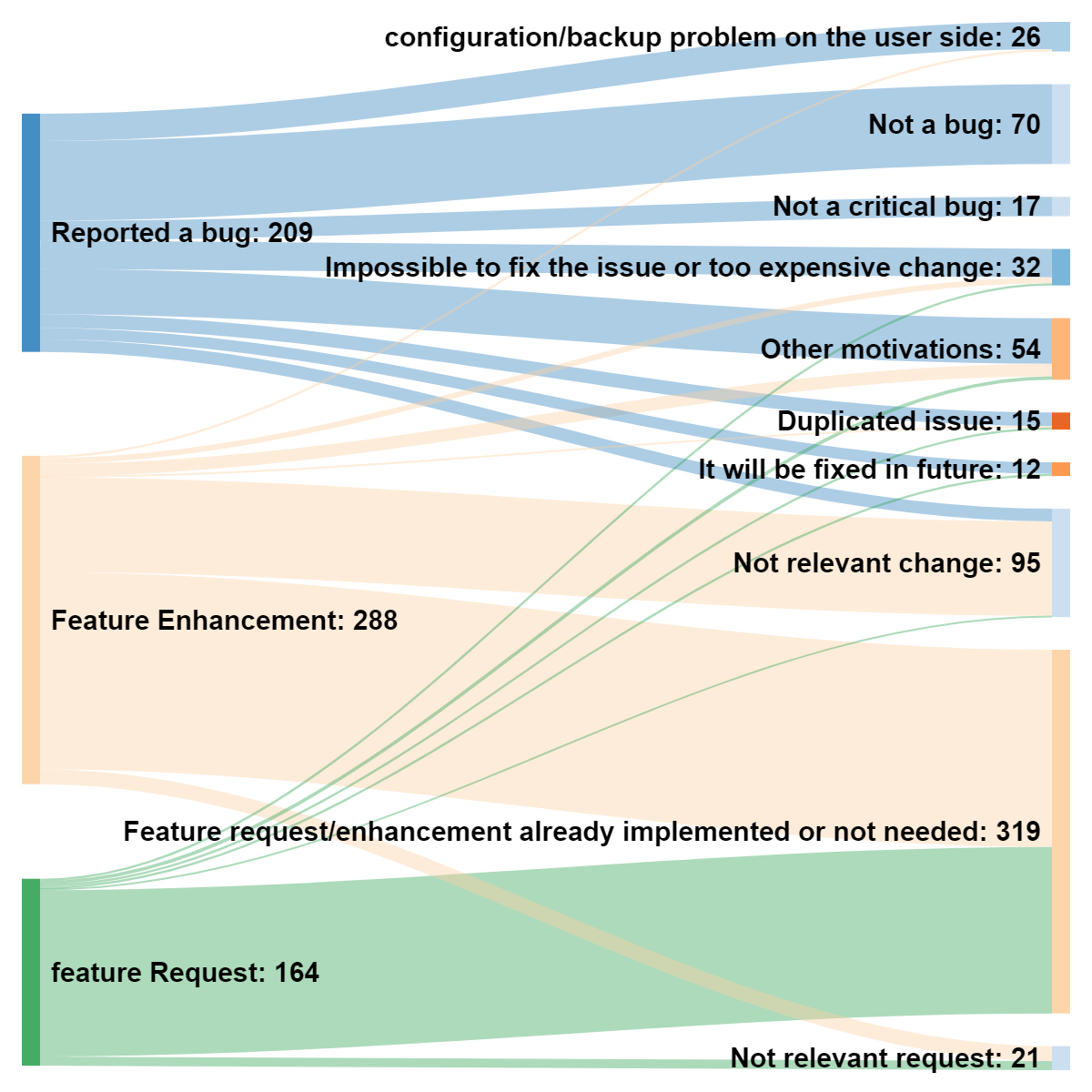} 
	\caption{\revise{Co-occurrences of motivations for issue opening and closing (only wontfix issues)}}
	\label{fig:coOccurr}
\end{figure*}

In Figure \ref{fig:coOccurr} we illustrate the frequency with which issues opened with  the most recurrent purposes in M$_{opening}$ (\ie \textit{Feature enhancement, Reported a bug} and \textit{Feature Request}) are closed with one of the motivations in M$_{closing}$ (see Table \ref{tab:label1}). \impr{In Figure \ref{fig:coOccurr} the thickness of the lines is proportional to the amount of issues opened with a specific reason (on the left) and closed with a specific M$_{closing}$ motivation (on the right).}
In particular, \revise{146} (\revise{89\%}) issues opened for requiring a new feature (\ie \textit{Feature Request}) have been closed with the  motivation \textit{Feature request/enhancement already implemented or not needed}, while \revise{173} issues (\revise{60.1\%}) requiring feature enhancements (\ie \textit{Feature Enhancement}) have been closed due to the same motivation. Moreover, 83 (\revise{28.8\%}) issues having the same purpose (\ie \textit{Feature Enhancement}) have been closed, since they proposed \textit{Not relevant changes}. Finally, issues reporting bugs (\ie  \textit{Reported a bug})  are mainly closed because (i) they do not signal actual bugs (\ie \textit{Not a bug}, \revise{33.5\%}),
(ii) the bugs reported are too expensive or impossible to fix (\ie \textit{Impossible to fix the issue or too expensive change}, \revise{12\%}), or 
(iii) \impr{the signaled defects mainly depend on configuration/backup problems on the user side (\ie \textit{unset backup or other configurations required on the user side for enabling the main functionalities of the project}, \revise{11.5\%})}.

\smallskip
\smallskip
{\centering
\fbox{\begin{minipage}{25em}
\small
\textbf{RQ$_1$ summary}: \emph{
Developers mainly tend to close issues (with the wontfix label) containing erroneous reports, or requesting features (or changes) that are not relevant or not needed.} 
\end{minipage}}
}
\smallskip

\subsection{\textbf{RQ$_2$: Factors Related with the Wontfix Issues Resolution Time}}
\label{sec:rq2_results}
\begin{table}[t!]
   \begin{center}
       \small
       \caption{\major{Summary of wontfix issues characteristics}}
       \label{tab:metricsAllData}
       \begin{tabular}{l r r} \toprule
	     \textbf{Metric} & \textbf{Median} & \textbf{Average} \\ \toprule
            nCommentsT & 3.00 & 4.43 \\
            nActorsT & 2.00 & 2.34 \\
            maxAuthorPercentage & 0.50 & 0.63 \\
            minorAuthors & 0.00 & 0.82 \\
            majorAuthors & 2.00 & 1.54 \\
            timeToCloseIssue & 42.60 & 153.59 \\
            timeToDiscussIssue & 76.05 & 201.33 \\
            DescriptionLength & 470.00 & 847.49 \\
            meanCommentSize & 380.00 & 495.68 \\ \bottomrule
	\end{tabular}
 \end{center} 
\end{table}

\begin{table*}[h!]
   \begin{center}
       \caption{\revise{Hypothesis testing results: All (A), Bug (B), \major{Not suitable (NS)}, Feature request/enhancement (FR), Change (C)}}
       \label{tab:hypMetrics}
       \resizebox{0.9\textwidth}{!}{
       \begin{tabular}{l c c c c c c c c c c} \toprule
	     \textbf{Metric} & \textbf{A-B} & \textbf{A-NS}  &
	     \textbf{A-FR} & \textbf{A-C} & \textbf{B-NS} & \textbf{B-FR} &
	     \textbf{B-C} & \textbf{NS-FR} & \textbf{NS-C} & \textbf{FR-C} \\ \toprule
nCommentsT	& $p > 0.1$ & $p > 0.1$ & $p < 0.05$ & $p < 0.05$ & $p > 0.1$ & $p < 0.05$ & $p > 0.1$ & $p < 0.05$ & $p < 0.05$ & $p < 0.05$ \\ 
nActorsT	& $p > 0.1$ & $p > 0.1$ & $p > 0.1$ & $p < 0.05$ & $p > 0.1$ & $p > 0.1$ & $p < 0.05$ & $p > 0.1$ & $p < 0.05$ & $p < 0.05$ \\ 
maxAuthorPercentage & $p > 0.1$ & $p > 0.1$ & $p < 0.05$ & $p < 0.05$ & $p > 0.1$ & $p > 0.1$ & $p < 0.05$ & $p > 0.1$ & $p < 0.05$ & $p < 0.05$ \\
minorAuthors & $p > 0.1$ & $p > 0.1$ & \major{$p > 0.1$} & $p < 0.05$ & $p > 0.1$ & $p > 0.1$ & \major{$0.05 \leq p \leq 0.1$} & $p > 0.1$ & $p < 0.05$ & $p < 0.05$ \\
majorAuthors & $p > 0.1$ & $p > 0.1$ & $p > 0.1$  &  $p > 0.1$ &  $p > 0.1$ &	$p > 0.1$ & $p > 0.1$ & $p > 0.1$ & $p > 0.1$ & $p > 0.1$ \\
timeToCloseIssue & $p > 0.1$ & $p < 0.05$ & \major{$p > 0.1$} & $p > 0.1$ & \major{$p > 0.1$} & $p > 0.1$ & $p > 0.1$ & $p < 0.05$ & \major{$p > 0.1$} & $p > 0.1$ \\
timeToDiscussIssue & $p > 0.1$ & \major{$p > 0.1$} & $p > 0.1$ & $p > 0.1$ & $p > 0.1$ & $p > 0.1$ & $p > 0.1$ & \major{$0.05 \leq p \leq 0.1$} & $p > 0.1$ & $p > 0.1$ \\
DescriptionLength & $p > 0.1$ & $p > 0.1$ & \major{$p > 0.1$} & $p < 0.05$ & $p > 0.1$ & $p > 0.1$ & \major{$p > 0.1$} & \major{$p > 0.1$} & $p > 0.1$ & $p < 0.05$ \\
meanCommentSize	 & $p > 0.1$ & $p > 0.1$ & $p > 0.1$ & \major{$p > 0.1$} & $p > 0.1$ & $p > 0.1$ & \major{$p > 0.1$} & $p > 0.1$ & $p > 0.1$ & \major{$0.05 \leq p \leq 0.1$} \\ \bottomrule
	\end{tabular}
	}
  \end{center} 
\end{table*}

\begin{figure}
\centering 
	\includegraphics[scale=0.55]{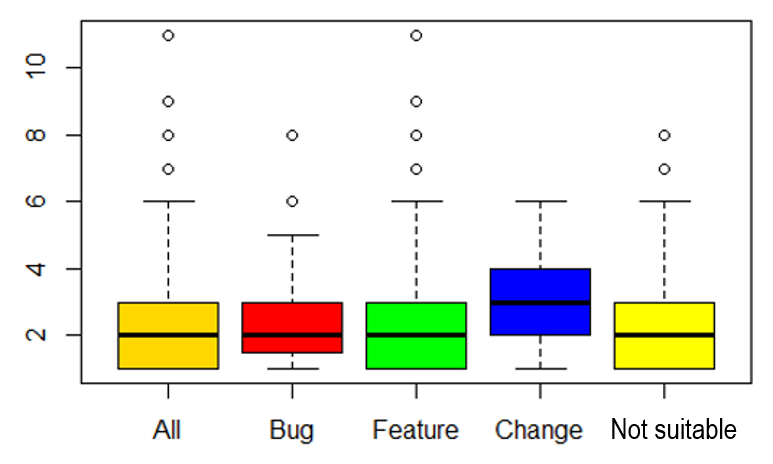} 
	\caption{\revise{Distributions of \textit{nActorsT} for the M$_{closing}$ categories, 
	All (median $=$ 2), Bug (median~$=$~2), Feature (median~$=$~2), Change (median~$=$~3), \major{Not suitable} (median~$=$~2)}}
	\label{fig:actors_bplot}
\end{figure}
\begin{figure}
\centering 
	\includegraphics[scale=0.55]{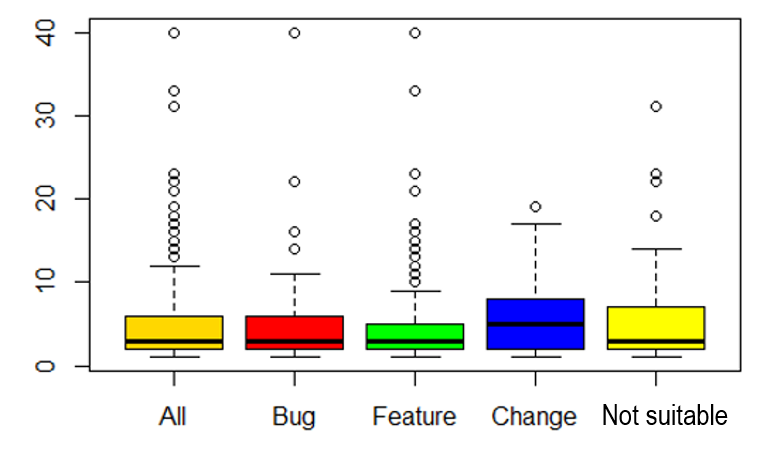} 
	\caption{\revise{Distributions of \textit{nCommentsT} for the M$_{closing}$ categories, 
	All (median~$=$~3), Bug (median~$=$~\revise{3}), Feature (median~$=$~\revise{3}), Change (median~$=$~5), \major{Not suitable} (median~$=$~3)}}
	\label{fig:comments_bplot}
\end{figure}
\begin{figure}
\centering 
	\includegraphics[scale=0.55]{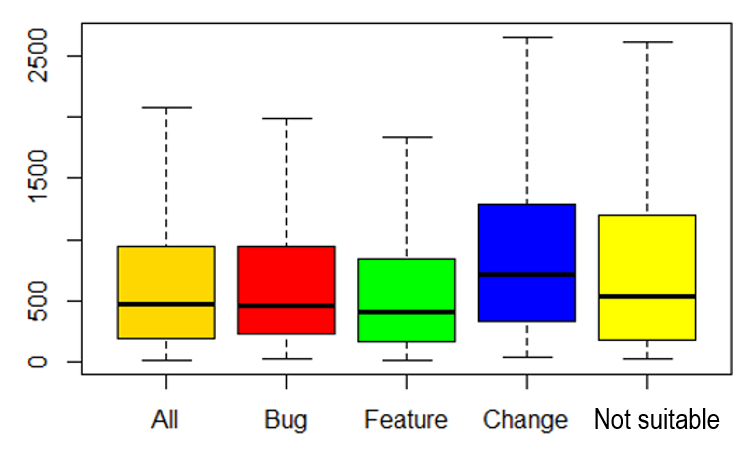} 
	\caption{\revise{Distributions of  \textit{descriptionLength} for the M$_{closing}$ categories, 
	All (median~$=$~\revise{470}), Bug (median~$=$~\revise{459}), Feature (median~$=$~\revise{408.5}), Change (median~$=$~\revise{720}), \major{Not suitable} (median~$=$~\revise{535.5})}}
	\label{fig:description_bplot}
	\vspace{-2mm}
\end{figure}
\begin{figure}
\centering 
	\includegraphics[scale=0.55]{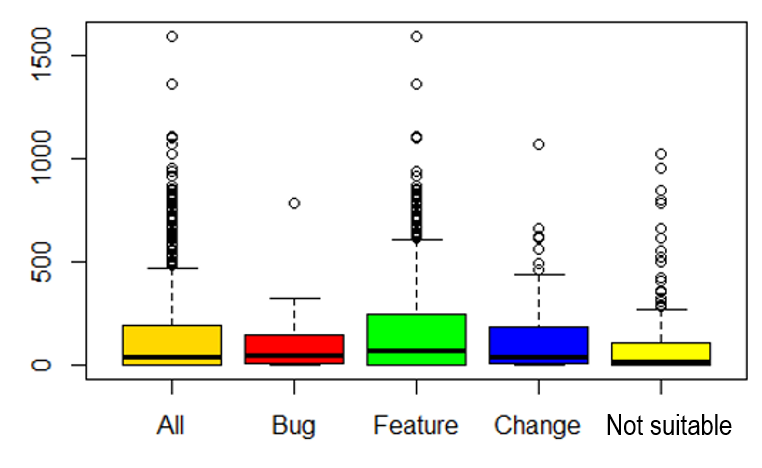} 
	\caption{\revise{Distributions of  \textit{timeToCloseIssue} for the M$_{closing}$ categories,
	All (median~$= \revise{42.60}$), Bug (median~$= \revise{52.06}$), Feature (median~$= \revise{73.59}$), Change (median~$= \revise{39.19}$), \major{Not suitable} (median~$= \revise{15.67}$)}}
	\label{fig:timeClose_bplot}
\end{figure}

\major{As reported in Table \ref{tab:metricsAllData}, wontfix issues are mainly discussed among limited numbers of major actors and such discussions encompass 4.43 comments, on average. As anticipated, wontfix issues are closed very long time after their opening (\ie about five months on average) and continue to be discussed even after their closing.} 

In order to investigate differences in the different types of wontfix issues, we study the extent to which the collected metrics vary across the specific $M_{closing}$ categories, and discuss the most interesting peculiarities. 
\impr{More specifically, to study  the differences occurring between the different kinds of wontfix issues, and verify whether the observed differences are statistically relevant, we tested the following null hypothesis:}

\begin{flushleft}
\impr{\emph{$H_{0}$: The distributions of values of the metric m for the populations of wontfix issues of the types t$_i$ and t$_j$ are equal}} 
\end{flushleft}

$\forall m \in$ \{\textit{descriptionLength,  maxAuthorPercentage,  majorAuthors, meanCommentSize, minorAuthors, nActorsT, nCommentsT, timeToCloseIssue, timeToDiscussIssue}\}, 
$\forall t_i, t_j \in$ \{\textit{All, Bug, \major{Not suitable}, Feature request/ enhancement, Change}\} and $i \neq j$.

$H_{0}$ has been tested with Mann-Whitney test and the \textit{p-value} was fixed to .05. Table \ref{tab:hypMetrics} reports the results of the Mann-Whitney test. \impr{This investigation is aimed at understanding whether specific types of wontfix issues exhibit more unproductive (or longer) discussions, before their closing.} \impr{In particular, for each metric (on the rows) and each pair of wontfix issue types (on the columns) Table \ref{tab:hypMetrics} reports the p-value obtained when testing the null hypothesis H$_0$ through the Mann-Withney U test.} \major{The Benjamini-Hochberg correction procedure~\cite{benjamini1995controlling} has been adopted to adjust p-values, since multiple comparisons are performed simultaneously.}

\begin{figure*}
\centering 
	\includegraphics[width=0.95\textwidth]{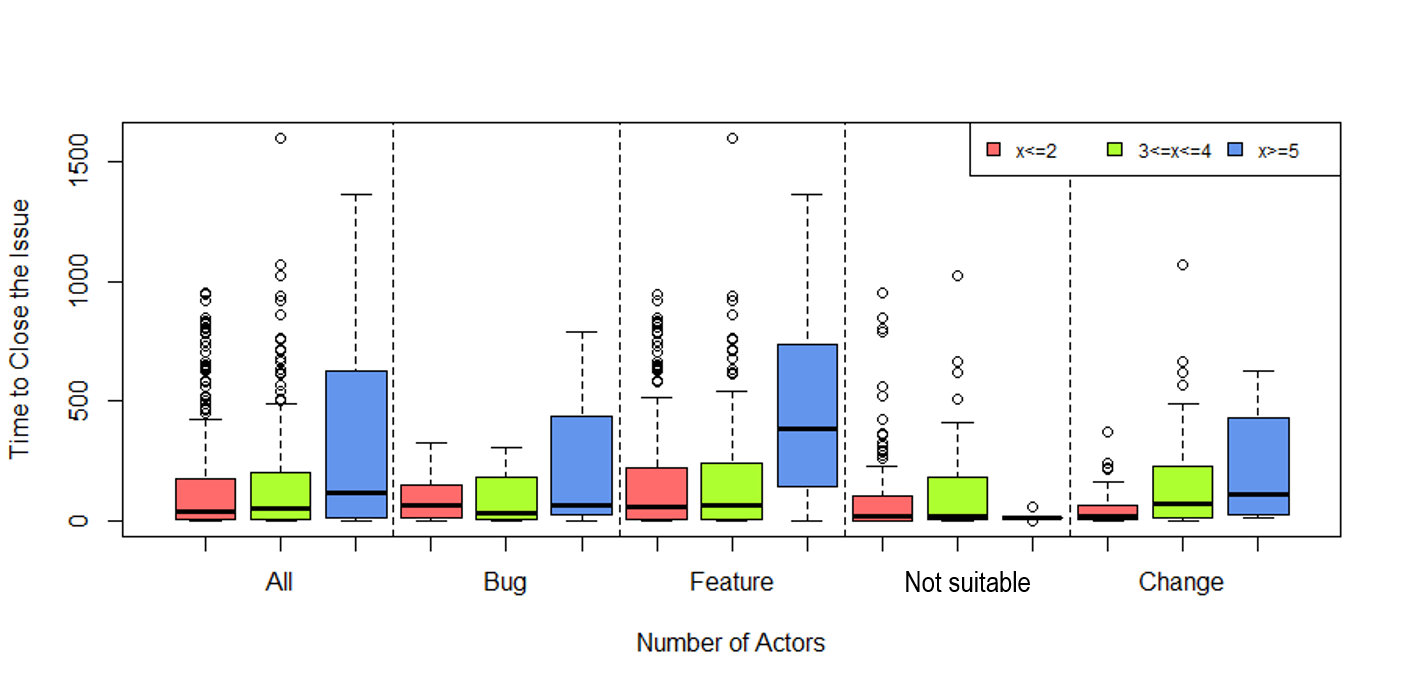} 
	\caption{\revise{Distributions of  \textit{timeToCloseIssue} for issues (labeled as wontfix) having different number of actors ($x$): $x \leq 2$, $3 \leq x \leq 4$, $x \geq 5$}}
	\label{fig:actorsTime_bplots}
\end{figure*}

Figures \ref{fig:actors_bplot}, \ref{fig:comments_bplot}, \ref{fig:timeClose_bplot} and  \ref{fig:description_bplot} report the respective distributions of \textit{nActorsT, nCommentsT, timeToCloseIssue} and \textit{descriptionLength} obtained for the overall issues in our dataset (\ie \textit{All}), as well as the various $M_{closing}$ categories. 
Change requests (\ie \textit{Change}) (i) are usually described through longer texts (see Figure \ref{fig:description_bplot}), (ii) require to be discussed between a greater number of actors (as illustrated in Figure \ref{fig:actors_bplot}) and, consequently, (iii) the related issue discussions comprise a greater amount of comments (see Figure \ref{fig:comments_bplot}), than other kinds of issues. 
As shown in Figure \ref{fig:timeClose_bplot}, \major{not suitable} reports (\ie \textit{\major{Not suitable}}) are closed much faster than the other types of issues: 50\% of issues of this type are closed in less than \revise{16} days, while the 50\% of the issues belonging to the other categories require more than \revise{39} days to be closed.  Probably, this is due to the fact that developers are more resolute in closing the issue, once verified that the signaled defect is not \major{actually suited to be addressed}. 

On the contrary, \textit{Feature requests/enhancement} issue types usually require more time to be closed (the median value of \textit{timeToCloseIssue} obtained for issues of this type is \revise{73.59} days), probably because developers have greater uncertainty on deciding if the required improvements could be in line with the project's purposes. 
In general, the number of participants discussing the issues may influence, with statistical evidence (see Table~\ref{tab:hypTimeTable}), the time required to close a wontfix issue, while for the other collected factors we do not observe significant relationships. Specifically, as illustrated in Figure~\ref{fig:actorsTime_bplots}, when the numbers of actors participating in the issue discussions concerning  wontfix issues of the \textit{Feature request/enhancement} and \textit{Change} types increase, we observe a longer \textit{timeToCloseIssue}, while for wontfix issues of the \textit{Bug} and \textit{\major{Not suitable}} types, no statistically significant differences between the different subsets are revealed. \rev{It is worth noticing that we verified whether significant relationships exist between the other metrics and the \textit{timeToCloseIssue}, by using similar analyses. However, such analyses did not produce noteworthy results.} 

\begin{table}
   \begin{center}
       \caption{\revise{Hypothesis testing results for issues having different number of actors ($x$)}}
       \label{tab:hypTimeTable}
              \resizebox{0.5\textwidth}{!}{
       \begin{tabular}{l c c c c c} \toprule
	     \textbf{Test} & \textbf{All} & \textbf{Bug}  &
	     \textbf{Feature} & \textbf{\major{Not suitable}} & \textbf{Change} \\ \toprule
	    $(x \leq 2)$ \textit{vs} $(3 \leq x \leq 4)$ & $p < 0.05$ & $p > 0.1$ & $p > 0.1$ & $p > 0.1$ & $p < 0.05$ \\
	    $(x \leq 2)$ \textit{vs} $(x \geq 5)$ & $p < 0.05$ & $p > 0.1$ & $p < 0.05$ & $p > 0.1$ & $p < 0.05$ \\
	    $(3 \leq x \leq 4)$ \textit{vs} $(x \geq 5)$ & $p < 0.05$ & $p > 0.1$ & $p < 0.05$ & $p > 0.1$ & $p > 0.1$ \\ \bottomrule
	\end{tabular}
	}
 \end{center} 
\end{table}
 
\major{In a study involving more than 4000 GitHub projects and about 1 million issues, Kikas \etal~\cite{Kikas:2016:UDC:2901739.2901751} found that the median lifetime of about 70\% of the investigated issues is 3.7 days. We observe that the median closing time for wontfix issues is about 11.5 times slower than the median lifetime of most issues investigated in prior work, confirming (i) our intuition that wontfix issues usually remain open for a longer time compared to other types of issues, and (ii) the need for early detection of issues that will probably remain unfixed.  Our study also partially confirms some of the findings reported in prior research~\cite{Kikas:2016:UDC:2901739.2901751}, showing that the number of different actors involved in issue discussion is related to the time to close the issue.}

\smallskip
{\centering
\fbox{\begin{minipage}{25em}
\small
\textbf{RQ$_2$ summary}: \emph{
On average, about five months are required to close issues that developers will label as wontfix. This time is mainly connected with (i) the issue type (issues indicating \major{not suitable} reports are closed much faster with respect to other kinds of issues), and (ii) the number of participants involved in the related discussions. Such discussions typically comprise less than 6 messages and involve a limited set of major actors. 
} 
\end{minipage}}
}
\smallskip

\subsection{\textbf{RQ$_3$: Automated Classification of Wontfix Issues}}
\label{sec:rq3_results}

As explained in Section \ref{sec:rq3_design}, we experimented with different ML techniques, namely (i) the probabilistic Naive Bayes classifier, (ii) SMO algorithm, and (iii) the J48 tree model. These ML models were trained on the training data (\ie \textbf{M$_{training}$}) and evaluated on the test  data (\ie \textbf{M$_{test}$}) illustrated in Section \ref{sec:rq3_design}. 
Table~\ref{tab:rq3Results} provides an overview of the main results obtained through the different ML algorithms. For completeness, in Table \ref{tab:rq3_NB}, Table \ref{tab:rq3_SMO} and Table \ref{tab:rq3_J48} we also provide the actual corresponding confusion matrices of all the three experimented ML models. 

\begin{table}
\centering
\revise{
\caption{Evaluation of Machine Learning Classifiers}
\label{tab:rq3Results}
\resizebox{0.85\linewidth}{!}{
\begin{tabular}{l|c|c|c}
	\toprule
	\textbf{ML Technique} & \textbf{Precision} & \textbf{Recall} & \textbf{F-Measure} \\ 
	\midrule
	Naive Bayes  & 0.731 & 0.698 & 0.708 \\
    SMO & 0.758 & 0.768 & 0.760  \\
    J48 & \textbf{0.896} & \textbf{0.897}  & \textbf{0.894}  \\
	\bottomrule
\end{tabular}
}
}
\vspace{-0.18cm}
\end{table}

The results in Table \ref{tab:rq3Results} highlight that the precision, recall, and \revise{F-measure values
are very positive for the J48 model, while we observe lower precision and F-measure results for the Naive Bayes and SMO models. 
Specifically, as reported in Table~\ref{tab:rq3Results} the J48 algorithm achieves the best classification performance, \ie values close to $0.90$ for precision, recall, and F-measure metrics. In the case of Naive Bayes and SMO, the values of precision, recall, and F-measure are lower than the ones achieved by the J48 model, with degradation in classification performance of more than 10\%.
The ML models perform a binary prediction (\ie \textit{wontfix} vs. \textit{non-wontfix}) relying on 14,720 textual features. The lower classification performance obtained by the Naive Bayes model 
could be due to the fact that, as reported by previous work on bug classification \cite{Antoniol:2008:BET:1463788.1463819}, 
``\emph{the naive Bayes classifier only exhibits a limited improvement when increasing  the number of features}'', while more complex machine learning models tend to achieve better classification performance, when the features' set grows up. }.

\begin{table}[h!]
\centering
\revise{
\caption{Naive Bayes: Confusion Matrix}
\label{tab:rq3_NB}
\resizebox{0.80\linewidth}{!}{
\begin{tabular}{|l|c|c|c|}\hline
		~ & ~ & \multicolumn{2}{c|}{\textbf{Predicted class}} \\		
	  ~ & ~ & \textbf{wontfix} & \textbf{non-wontfix} \\ \hline
	\textbf{Actual} & \textbf{wontfix} & 610 & 325\\\cline{2-4}
	\textbf{class} & \textbf{non-wontfix} & 632 & 1,598\\
	\hline
\end{tabular}
}
\vspace{-0.18cm}
}
\end{table}

\begin{table}[h!]
\centering
\revise{
\caption{SMO: Confusion Matrix}
\label{tab:rq3_SMO}
\resizebox{0.80\linewidth}{!}{
\begin{tabular}{|l|c|c|c|}	\hline
		~ & ~ & \multicolumn{2}{c|}{\textbf{Predicted class}} \\ 	
	      ~ & ~ & \textbf{wontfix} & \textbf{non-wontfix} \\ \hline
	\textbf{Actual} & \textbf{wontfix} & 482 & 453\\\cline{2-4}
	\textbf{class} & \textbf{non-wontfix} & 282 & 1,948\\
	\hline
\end{tabular}
}
}
\end{table}

   \begin{table}
\centering
\revise{
\caption{J48: Confusion Matrix}
\label{tab:rq3_J48}
\resizebox{0.80\linewidth}{!}{
\begin{tabular}{|l|c|c|c|}	\hline
		~ & ~ & \multicolumn{2}{c|}{\textbf{Predicted class}} \\
	 ~ & ~ & \textbf{wontfix} & \textbf{non-wontfix} \\ \hline
	\textbf{Actual} & \textbf{wontfix} & 702 & 233\\\cline{2-4}
				\textbf{class} & \textbf{non-wontfix} & 94 & 2,136\\
	\hline
\end{tabular}
}
}
\end{table}

The variability of the results can be easily explained by observing the confusion matrices of the three ML models, reported in Tables \ref{tab:rq3_NB}, \ref{tab:rq3_SMO} and  \ref{tab:rq3_J48}. \revise{For the J48 model the numbers of False Negatives (\textit{FN}) and False Positives (\textit{FP}) are relatively low, while in the case of the Naive Bayes and SMO ML strategies, the amount of misclassified instances is substantially higher.
Interestingly, the achieved results demonstrate that predicting whether an issue will be fixed or not (\ie will be marked as a wontifx) is possible with positive results, especially when considering a tree classifier. In addition, these results confirm our conjecture that terms occurring in the title and the description of issues posted on GitHub  are discriminant and relevant factors to consider for determining whether an issue will be fixed or not. 
}

\begin{figure}
\centering 
	\includegraphics[scale=0.39]{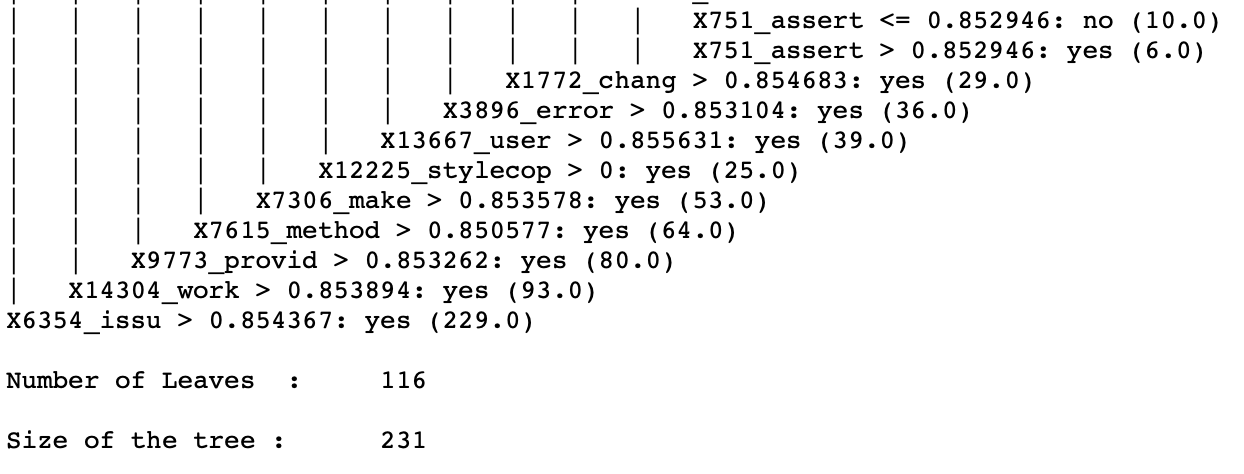} 
	\revise{
	\caption{The J48 tree model used for classification task of RQ3}
	\label{fig:j48-tree}
	}
\end{figure}

Our results are very encouraging, especially if we consider that we used just 50\% of our dataset to train the different ML algorithms and predicted on the remaining 50\%, \major{leading to equally balancing training and test sets}. Indeed, using a larger number of examples in the training set (\eg using 80\% of our dataset as training set and the remaining 20\% as test set, as done in traditional ML applications) is likely to result in higher performance. Thus, to check whether a larger number of points in the training set could lead to better results and mitigate concerns related to overfitting and selection bias, we repeated the classification experiment by using a 10-fold cross-validation strategy (\ie in each run of the 10-fold cross-validation the training set was composed by 90\% of items in the overall dataset). This analysis is also important to verify that the previously discussed results are not dependent on the specific data used for training the ML models. \rev{Indeed, one of the goals of using 10-fold cross-validation is to flag problems like overfitting or selection bias~\cite{Cawley:2010:OMS:1756006.1859921}.}

The results of 10-fold cross-validation are shown in Table \ref{tab:rq3Results2}. \revise{Such results confirm that the precision, recall, and F-measure values are very positive for J48, while lower precision, recall and F-measure results for the Naive Bayes and and SMO models are observed. More specifically, all the considered classifiers achieve slightly higher F-measure values (\ie with improvements ranging from 0.5\% to 4\%) than the results previously obtained (see Table \ref{tab:rq3Results}), confirming (i) the high performance of the J48 model in identifying the issues that will likely be not fixed by developers, and (ii) the inadequacy of the Naive Bayes technique when used for performing  the aforementioned classification task.}

\begin{table}
\centering
\revise{
\caption{Evaluation of Machine Learning Classifiers via 10-fold validation strategy}
\label{tab:rq3Results2}
\resizebox{0.85\linewidth}{!}{
\begin{tabular}{l|c|c|c}
	\toprule
	\textbf{ML Technique} & \textbf{Precision} & \textbf{Recall} & \textbf{F-Measure} \\ 
	\midrule
	Naive Bayes  & 0.724 &  0.710 & 0.715 \\
    SMO & 0.785 & 0.794  &  0.785 \\
    J48 & \textbf{0.938} & \textbf{0.936} & \textbf{0.934}  \\
	\bottomrule
\end{tabular}
}
}
\end{table}

\begin{table}[t]
\centering
\caption{\major{The 15 text features with the highest information gain scores}}
\label{tab:infoGain}
\small
\begin{tabular}{l|l} \toprule
\textbf{Score} & \textbf{Feature} \\ \toprule
 0.1467 & \texttt{X6354\_issu} \\
 0.1409 & \texttt{X14304\_work} \\
 0.1396 & \texttt{X9773\_provid} \\
 0.1150 & \texttt{X13840\_version} \\
 0.1058 & \texttt{X8222\_need} \\
 0.1050 & \texttt{X12642\_test} \\
 0.0986 & \texttt{X13272\_type} \\
 0.0966 & \texttt{X1772\_chang} \\
 0.0957 & \texttt{X12341\_support} \\
 0.0957 & \texttt{X3896\_error} \\
 0.0940 & \texttt{X2540\_creat} \\
 0.0910 & \texttt{X7306\_make} \\
 0.0907 & \texttt{X7615\_method} \\
 0.0899 & \texttt{X12433\_system} \\
 0.0859 & \texttt{X12182\_string} \\ \bottomrule
\end{tabular}
\end{table}

\revise{
To achieve more in-depth insights about the positive results obtained by the J48 model, we 
qualitatively/manually observed actual features that are selected by the J48 model to characterize/classify wontfix issues. 
From part of the complex J48 tree model in Figure \ref{fig:j48-tree} (see the full trained tree model in the RP\footnote{\url{https://bit.ly/2T2yjZG}}) used to perform the classification of issues, we can observe that the selected features concern in most cases textual features semantically linked to \textit{\textbf{requests for feature enhancement/addition}} such as \textit{make, change, provide}, etc. Interestingly, this result is in line with the finding of RQ$_1$, where we discovered that one of the major motivation for closing issues with the wontfix status (see Table \ref{tab:label1}) is the presence of not desired features (about 47\% of wontfix issues concern requests for features enhancement/addition).}
\major{To quantitatively corroborate this observation, we computed the information gain~\cite{quinlan1986induction} for all the text features leveraged by our model and ranked them according to their scores. In Table \ref{tab:infoGain}, the top 15 ranked features, along with the related information gain scores are reported.  
Looking at Table \ref{tab:infoGain}, it is easy to observe that many of the features with the highest scores are conceptually linked to requests for enhancement or feature additions (\eg \texttt{provide, need, change, support, create,} and \texttt{make}).
However, this could also represent a limitation of our model. Indeed, as shown in Table \ref{tab:rq3_J48}, while the J48 model is quite effective in recognizing issues of the \textit{non-wontfix} class, a higher likelihood of false negatives is observed for \textit{wontfix} issues (\ie the recall for this class is about 75\%). For this reason, we believe that further efforts and tunings could be aimed at reducing the false-negative rate for this class while keeping low the false-positive rate.
}


\bigskip
{\centering
\fbox{\begin{minipage}{25em}
	\revise{
\small
\textbf{RQ$_3$ summary}: \emph{
\begin{enumerate}
	\item Relying on a tree classifier (i.e., J48) it is possible to automatically detect issues that will be labeled as wontfix, with precision, recall, and F-measure values up to \textbf{$0.93$}. 
\item Consistently with the results of RQ$_1$ the experimented models select textual features semantically related to \textit{requests for features enhancement/addition} to classify wontfix issues.
\end{enumerate}
}
} 
\end{minipage}}
}
\bigskip

\section{Threats to Validity}
\label{sec:threats}
\textbf{Threats to construct validity}. In order to carry out our study, we measure different factors that could be not sufficient to model the whole issue handling process. As pointed out by Kalliamvakou \etal \cite{KalliamvakouMSR2014}, many active projects do not conduct all their software development activities in GitHub, and separate infrastructures (\eg mailing lists, forums, IRC channels, etc. \cite{PanichellaBPCA14}) could be used to support decision-making processes. This only represents a minor threat in our study, since most of the events of our interest (\eg opening and closing of the issues, label assignments, comments) are mostly recorded in the issue tracking system, as it is primarily used by development teams to track issue data. 

\textbf{Threats to internal validity}. Our results have been obtained by analyzing \textit{wontfix} issues having the \textit{closed} status assigned and such results could be misleading if a significant percentage of such issues will be reopened in the future. However, less than 9\% of non-fixed bugs tend to be reopened \cite{MiEASE2016} and one of the root causes for re-opening a bug report resolved as not fixed is due to the difficulty in reproducing the bug \cite{ZimmermannICSE2012}. It is worth to notice that in our manual inspection, the \textit{Difficult to fix or to replicate} and \textit{Not replicable bug} M$_{closing}$ motivations have been assigned only to 1.6\% and 0.1\% of issues in our sample, respectively. \revise{In addition, 281 out of 667 (42.13\%) wontfix issues encompassed in our manually analyzed sample are related to the \texttt{aspnet/Mvc} project. This could represent a threat to internal validity, as developers of this project could adopt similar criteria for deciding to avoid addressing a reported issue.}
\major{In our study we analyzed issues labeled as \textit{wontfix}. However, GitHub developers may indicate that a specific issue will be not resolved directly in the issue title (e.g., by using the \texttt{[WONTFIX]} prefix\footnote{See \url{https://github.com/waveform80/picamera/issues/657}}). To counteract this issue, we estimated the number of GitHub issues in C\# projects where the \texttt{WONTFIX} keyword appears in the issue title\footnote{\major{We used the following search string: ``is:issue \texttt{is:closed WONTFIX in:title language:C\#}"}}. Such a search returned a very limited number of issues (\ie $< 10$) and most of them (\ie 75\%) had also the \textit{wontfix} label assigned.}
\major{To avoid any bias in the potential evaluation of the performance of the experimented ML techniques, we adopted well-known information retrieval metrics, namely  precision, recall, and F-measure~\cite{BaezaYates:1999} and apply a cross-projects setting to train the ML models on data coming from different projects. However, since the information concerning all the issues in our dataset is used for the ML model construction, we can not exclude that characteristics related to more recent issues are leveraged to predict the resolution of previously submitted issues.}

\textbf{Threats to conclusion validity}. In our RQ$_2$, we analyze different issue clusters having different sizes in terms of issue numbers, and some of the differences we observed could be not significant. To mitigate this threat, we compared the values obtained for each cluster through the Mann-Whitney U test, widely adopted for similar purposes in the software engineering community, and discussed \rev{some of} the differences which resulted statistically significant (\textit{p-value} $< 0.05$). \major{Since multiple comparisons are performed simultaneously, the Benjamini-Hochberg correction procedure~\cite{benjamini1995controlling} has been adopted to adjust p-values and control the false discovery rate.}

\textbf{Threats to external validity} relate to the generalizability of our results. \revise{To mitigate this kind of threats our evaluation has been performed on a dataset containing 6,330 issues, extracted from 279 heterogeneous projects. Moreover, we manually analyzed 667 wontfix issues of 97 different projects. 
However, such sample could be not adequately representative of all the GitHub projects. 
All the considered wontfix issues (1,844) are related to projects developed using mainly the C\# programming language, and the average number of issues per projects tend to be quite skewed for some projects.  This may represent a threat to external validity, since such issues can present common characteristics that ease their identification. For these reasons, in the future, we aim at extending our investigation by evaluating wontfix issues in further projects developed through different programming languages. However, on a positive side, the classifier demonstrated to achieve high performance in identifying wontfix issues, even when not trained on issues related to any specific project, thus this training strategy allows the classifier to be more easily used on projects, different from the ones used in our experimentation, without the need for re-training it}.
\rev{On the other hand, the high classification performance achieved by some ML models could depend on the fact that many wontfix issues encompassed in our dataset concern \textit{requests for features enhancement/addition}. Indeed, the qualitative analysis performed in Section \ref{sec:rq3_results} highlighted that ML models leverage syntactical features semantically related to this kind of requests to perform the classification. Thus, it is not clear if similar results could be obtained on more balanced datasets, in which the different types of wontfix issues are equally represented.}

\section{Conclusions}
\label{sec:conclusions}
Software maintenance and evolution activities represent crucial tasks of any successful software project, and issues reported by users are a valuable source of information for developers interested in improving their systems. However, developers spend significant time handling issue reports and user requests. To support developers during issue handling processes, researchers conceived effective solutions aimed at prioritizing requested changes, as well as detecting potential issue misclassifications or duplications. 
However, few prior studies explored the nature of wontfix issues, and none of these studies proposed approaches to automatically determine whether an issue will be marked as \textit{wontfix}. We argue that a timely identification of issues that are likely to be not addressed, could help 
(i) project managers allocating resources, (ii) developers focusing their attention on the issues that will be actually addressed, and (iii) customers knowing early if their requirements would be satisfied~\cite{DBLP:journals/jss/Ramirez-MoraOG20}.
\revise{To this aim, in this paper, by collecting more than  6,000 issues extracted from the history
of 279 GitHub projects, we (i) analyzed the common characteristics of wontfix issues, and (ii) proposed an approach leveraging textual analysis and machine learning techniques to predict whether an issue will be resolved as a wontfix. Results of our study show that developers
mainly tend to close issues (with the wontfix label) containing erroneous reports, or requesting features (or changes) that are not relevant or out of the purposes of projects (RQ$_1$). However, developers take a significant amount of time (about five months, on average) to decide whether an issue should be labeled as a wontfix. This time is mainly connected with (i) the issue type (issues containing erroneous bug reports, are closed much faster), and (ii) the number of participants involved in the related discussions (RQ$_2$). 
Last but not least, the proposed approach proved to be accurate (with a F-measure up to 93\%) in identifying issues that will be likely labeled as wontfix (RQ$_3$).}

This study helps to better comprehend the issue management dynamics in open source communities. As a future work, we aim at investigating whether different projects tend to have different  wontfix characteristics (due to different issue management processes), and the extent to which the automated identification of wontifx issues may impact the results produced by issue prioritization approaches. \impr{In addition, in the future, we aim at comparing how issues with/without \textit{wontfix} label perform each other, in order to investigate how the presence of \textit{wontfix} issues may affect the overall issue management process.}  
\rev{We intend to also study further wontfix factors useful to automatically identify/predict the actual potential motivations (as it could be useful information for developers) behind an issue that will be closed as a wontfix. Moreover, we plan to compare the results of our approach with machine learning approaches successfully used in the same context~\cite{KallisICSME2019, KALLIS2021102598} and involving other types of labels.
Finally, we plan to  investigate the usage of historical analysis to provide orthogonal/complementary information that could be combined with the adopted textual features}.

Differently from issue driven development, in pull-based development developers use branches to make the desired changes independently, and then create a pull request to ask merging their changes into the main repository~\cite{DBLP:conf/icse/GousiosSB16}. The integrators (usually the members of the project's core team) are asked to reply to such request, evaluating the quality of the contributions, and eventually merging or rejecting the changes~\cite{DBLP:conf/icse/GousiosZSD15}. Manually identifying high-quality and desirable pull requests may be challenging~\cite{DBLP:conf/internetware/YangZZFYW17}, especially for popular projects, where tens of pull requests are daily submitted~\cite{DBLP:conf/icsm/YuWYL14, Azeem:2020}. In the future we plan to verify whether the reasons behind the rejection of specific kinds of pull requests are similar to the ones that have been identified for wontfix issues, with the purpose of better comprehending the team behaviors when managing external requests.

\section*{Acknowledgment}
Sebastiano Panichella gratefully acknowledges the Horizon 2020 (EU Commission) support for the project \textit{COSMOS} (DevOps for Complex Cyber-physical Systems), Project No. 957254-COSMOS.


\balance
\bibliographystyle{elsarticle-num}
\bibliography{biblio}

\end{document}